\documentclass{article}%
\usepackage[english]{babel}
\usepackage{amsmath}
\usepackage{amsfonts}
\usepackage{amssymb}

\newcommand{\be}{\begin{equation}}
\newcommand{\ee}{\end{equation}}

\def\la{\mathrel{\mathpalette\fun <}}
\def\ga{\mathrel{\mathpalette\fun >}}
\def\fun#1#2{\lower3.6pt\vbox{\baselineskip0pt\lineskip.9pt
\ialign{$\mathsurround=0pt#1\hfil##\hfil$\crcr#2\crcr\sim\crcr}}}

\begin{document}

\title{From super-charged nuclei to massive nuclear density cores}


\author{Vladimir Popov \\
{\it\small Institute of Theoretical and Experimental Physics 117218 Moscow, Russia}}

\date{}

\maketitle

\begin{abstract}
Due to $e^+e^-$-pair production in the field of supercritical $(Z \gg Z_{cr}\approx 170
$) nucleus an electron shell, created out of the vacuum, is formed. The distribution of
the vacuum charge in this shell has been determined for super-charged nuclei $Ze^3 \ga 1$
within the framework of the Thomas-Fermi equation generalized to the relativistic case.
For $Ze^3 \gg 1$ the electron shell penetrates inside  the nucleus and almost completely
screens its charge. Inside such nucleus the potential takes a constant value equal to
$V_0=-(3\pi^2 n_p)^{1/3} \sim -2m_{\pi}c^2$,  and super-charged nucleus represents  an
electrically neutral plasma consisting of $e,p$ and $n$. Near the edge of the nucleus a
transition layer exists with a width $\lambda \approx  \alpha^{-1/2} \hbar/m_{\pi} c\sim
15$ fm, which is independent of $Z~~ (\hbar/m_{\pi} c \ll \lambda \ll \hbar/m_e c)$. The
electric field and surface charge are concentrated  in this layer. These results,
obtained earlier for hypothetical  superheavy nuclei  with  $Z \sim A/2\la 10^4 \div
10^6$, are extrapolated to massive nuclear density cores having a mass number $A \approx
(m_{Planck}/m_n)\sim 10^{57}$. The problem of the gravitational and electrodynamical
stability of such objects is considered. It is shown that for $A \ga 0.04
(Z/A)^{1/2}(m_{Planck}/m_n)^3$ the Coulomb repulsion of protons, screened by relativistic
electrons, can be balanced by  gravitational forces. The overcritical electric fields
$E\sim m^2_{\pi} c^3/e\hbar$ are present in the narrow transition layer near the core surface.
\end{abstract}

The Dirac equation for an electron in the field  of a point-like electric charge $Ze$
loses its sense for $Z > 137$,  since the energies $\varepsilon_n$ of the bound  states
$n s_{1/2}$ and $np_{1/2}$ become complex [1]-[3]. For instance, in the case of the
lowest energy levels one has
\be \varepsilon(1 s_{1/2}) = m_e c^2 \sqrt{1-\zeta^2} - \mbox{for the ground
state},\label{1}\ee

$$ \varepsilon (2s_{1/2}) =\varepsilon (2p_{1/2}) = m_e c^2
\sqrt{ \frac{1+ \sqrt{1-\zeta^2}}{2} },\eqno(1')$$

$$\varepsilon(2p_{3/2}) = m_e c^2\sqrt{1-\zeta^2/4}, ...,$$ where $0 < \zeta \equiv Z\alpha <
1$, $\alpha = e^2/\hbar c=1/137$,  $m_e$ is the electron mass and the potential is
assumed to be $V(r) =-\zeta/r$, $0< r <\infty$. Analogous singularities at $\zeta =1$
appear for all $ns_{1/2}$ and $np_{1/2}$ states:
$$ \varepsilon_n / m_e c^2 =\frac{n-1 +\sqrt{1- \zeta^2}}{[~N^2 +
2(n-1)\sqrt{1-\zeta^2}~]^{1/2}} = $$

$$ =\frac{n-1}{N} + \frac{1}{N^3}\sqrt{1-\zeta^2} + \frac{(n-1)(N^2-3)}{2N^5}(1-\zeta^2)+$$
\be+ O ((1-\zeta^2)^{3/2}),~~~\zeta \to 1, \label{2}\ee where $N=\sqrt{n^2 - 2n +2}$ and
 $\varepsilon_n =m_e c^2\sqrt{1-N^{-2}}$ at $\zeta=1$. In particular, for the case of
 highly-excited, $n\gg  1$, states
 $$ \varepsilon/m_e c^2=1 -\frac{\zeta^2}{2n^2}-\frac{\zeta^4}{n^3(1+\sqrt{1-\zeta^2)}} +
 ...,~~0 <\zeta < 1.\eqno(2')$$

The appearance of complex energies $\varepsilon_n(\zeta)$ at $\zeta > 1$ contradicts to
unitarity and hermiticity of the Dirac Hamiltonian, so an immediate analytical
contituation of the previous formulae to $Z > 137$ region is impossible. Analogous
singularities exist for  other physical quantities, for example, for the mean radius and
the magnetic moment of the ground state,
$$ \langle r \rangle = \frac{1+2\sqrt{1-\zeta^2}}{2\zeta} \cdot l_C,~~~\mu
=  \frac{1}{3} (1+2\sqrt{1-\zeta^2})\mu_B,\eqno{(1'')}$$ where $l_C =\hbar/m_e c =\alpha
a_B$ and $\mu_B =e\hbar/2 m_e c$.

Many aspects of Quantum Electrodynamics of strong fields  are considered in refs.[5-29],
including  the relativistic Coulomb problem with $Z > 137$ [5,6,11-15], the critical
nuclear charge $Z_{cr}$ [5,11,15], vacuum polarization and superbound electrons in the
lower continuum at $Z
> Z_{cr}$ [10,13,17], spontaneous production of positrons at $Z
> Z_{cr}$ [6,8,12] and in collisions
of two heavy nuclei with united charge $Z_1+Z_2 > Z_{cr}$ [10,13,14], the Thomas-Fermi
approach for super-charged nuclei [7,18], $e^+e^-$-pair creation from vacuum in strong
electric field and by intense  laser radiation [22-27], etc. For further details see the
reviews [9,10,13,21,28].

The difficulty \footnote{Similar situation takes place in other problems with the
so-called ``fall down to the center'' in quantum mechanics, see \S~ 35 in ref.[30]. For
solutions of the Dirac equation with a point-like Coulomb potential ``fall down to the
center'' begins at $\zeta=j+1/2$, where  $j=1/2, 3/2, 5/2, ...$ is the total angular
momentum of the electron state [10,31].} arising at $\zeta \geq 1$, which sometimes is
called ``$Z=137$ catastrophe'' [28], is removed
 if one takes into account
the finite size of nucleus [4]
\be V(r) = -\frac{Z\alpha}{r} f (r/r_N), \label{3}\ee where the cut-off function $f$
depends on the distribution of electric charge inside the nucleus. For example, the
function
$$ f(x) = \left\{ \begin{array}{ll}
x(3-x^2)/2, & 0 < x < 1\\ & \\ 1, & x \equiv r/r_N > 1\end{array} \right. \eqno(3')$$
corresponds to uniform  volume density of  electric charge $Ze$ and is frequently used in
calculations. Here the potential at the centre of nucleus is finite, $V(0)=-1.5
Z\alpha/r_N$.

When a finite radius $r_N >0$ is introduced, the ground level $1 s_{1/2}$ is going down
monotonously while $Z$ increasing  and reaches the boundary of the lower continuum
$\varepsilon = -m_e c^2$ for $Z = Z_{cr}\approx 170$ [5,11,15]. It can be shown that the
asymptotic expressions for $1 s_{1/2}$ energy are [12,13]
\be \frac{\varepsilon(1 s_{1/2})}{ m_e c^2} = \left\{ \begin{array}{ll} \sqrt{1-\zeta^2}~
\mbox{coth}~(\Lambda\sqrt{1-\zeta^2}), & 0 < \zeta \leq 1, \\ &\\
\sqrt{\zeta^2-1}~ \mbox{ctg}~(\Lambda\sqrt{\zeta^2-1}), & \zeta \geq
1,\end{array}\right.\ee where $\zeta = Z\alpha$, $\Lambda = \ln (l_C/r_N) \gg 1$ is a
logarithmic parameter in the problem considered and $l_C =\hbar/m_e c = 386$ fm is the
Compton wave length for electron. Eqs.(4) explicitly show that $\zeta=1$ is not a
singular point for the energy $\varepsilon(\zeta)$, on the contrary to the case of a
point-like charge, and energy levels $\varepsilon(n,j)$ of the bound states $1s, 2p, 2s,
...$ smoothly continue to drop into the lower continuum as $Z$ increases, until at
$Z=Z_{cr}(n,j)$ its energy reaches the boundary of the lower continuum. Numerical values
of the ``critical nuclear charge'' $Z_{cr}(n,j)$ were obtained by different calculation
methods from the equation $\varepsilon(n,j)=-m_e c^2$, see [5,10,11-13] and references
therein. A simple asymptotic formula for  $Z_{cr}$ follows from eq.(4) and analogous
equations for $2p,2s$, ... states:
\be \tilde{Z}_{cr}  \alpha = 1 + \frac{n^2_r \pi^2}{2\Lambda(\Lambda + c_{nj})} + O
(\Lambda^{-4}), \label{5}\ee where $n_r = n$ for $ns$-states, $n_r = n-1$ for
$np_{1/2}$-states, $n$ = 1,2,3, ... is the principal quantum number, $c_{nj}=2n$ for
$ns$-states, $c_{nj}=\sqrt{2}-1$ for $2p_{1/2}$-state and the condition $\Lambda \gg n$
was assumed. As can be seen from the following Table, the approximation (5) is rather
good for the lowest levels of the electron spectrum, though the expansion parameter
$\Lambda \sim 3.7$ is not very  large.

\begin{center}

{\bf Critical nuclear charge for the low-lying  states of electron spectrum}

\vspace{5mm}

\begin{tabular}{|l|llllll|}\hline
&&&&&&\\ Atomic  & $Z_{cr}^{(0)}$  & $Z_{cr}$ & $\tilde{Z}_{cr}$ &
$\zeta_{cr}$ & $r_N, fm$ & $\Lambda$ \\ state &&&&&& \\ &&&&&& \\\hline &&&&&&\\
$1s_{1/2}$ & 168.8 & 172
& 169 & 1.255 & 9.14 & 3.74 \\ &&&&&& \\
$2p_{1/2}$ & 181.3 &  185 & 181 & 1.350 & 9.33 & 3.72 \\ &&&&&& \\ $2s_{1/2}$ &
232 & 239 & 232 & 1.745 & 10.1 & 3.64 \\ &&&&&& \\  $3p_{1/2}$ & 254 & 263 &
-- & 1.920
 & 10.5 & 3.60 \\ &&&&&& \\
\hline
\end{tabular}

\end{center}
 {\bf Footnote to the Table:} the values of $Z^{(0)}_{cr}$ correspond to the ``naked
 nucleus'' with the cut-off model (3$'$), $Z_{cr}$ are calculated with account of screening
 of the Coulomb field $V(r)=-\zeta/r$
 by outward electrons (except of the $K$-shell, which is supposed to
 be ionized), $\zeta_{cr} = Z_{cr}/137$, and the values $\tilde{Z}_{cr}$ are calculated
by the asymptotic formula (5).

\bigskip

Note that the electric field $E(r)$ near surface of a heavy nucleus, $r\approx r_N = 1.2
A^{1/3}$ fm, is much larger than the ``critical'' or Schwinger field in QED [22]:
\be E_{cr} = m^2_e c^3/e\hbar = 1.32 \cdot 10^{16} ~\mbox{V/cm},\label{5}\ee and the
Coulomb field of a heavy nucleus is equal to
\be E(r) = Ze/r^2,~~~r > r_N; ~~~E(r_N)/E_{cr} = Z\alpha~(l_C/r_N)^2.\label{7}\ee So,
$E(r_N)$ $\approx 1800~ E_{cr}$ for $U$ nucleus ($Z=92,  A=238, r_N = 7.45$ fm) and
$E(r_N)\approx$
\\ $ 2200$  $E_{cr}$ for $Z=172$ ($A=2.6~Z,~r_N =9.2$ fm). However, the static supercritical
field $E(r)$ is strongly inhomogeneous and exists only in a small space region near
$r_N$, therefore no $e^+e^-$ pairs can be produced by this field if $Z < Z_{cr}$. As is
seen from eq.(7), the Coulomb field is larger than $E_{cr}$ only at distances
\footnote{Note that $E(l_C)=E_{cr}$ if $Z=\alpha^{-1}=137$, and $E(r) > E_{cr}$  at $r <
\sqrt{Z\alpha}~l_C$.} $r \la l_C$.

The same is true also for the field (17) of the massive nuclear density core, where

\be E(r) \approx E_{max}\cdot \xi^{-2} \sim E_{max}  \biggl ( \frac{\lambda}{r-R_c}\biggr
)^2,~~~\xi \gg 1,\ee the parameters $\xi$ and $\lambda$ are defined in Eq.(14) below and
$E(r)
> E_{cr}$ only near the core radius,
\be r -R_c \la \lambda \sqrt{E_{max}/E_{cr}} \sim \alpha^{-1/4}l_C.\ee Therefore the
well-known formula [22] for pair production probability in homogeneous electrostatic
field, $w \propto (E/E_{cr})^2\cdot \exp (-\pi E_{cr}/E)$, is not applicable in these
cases.

For $Z > Z_{cr}$ the vacuum becomes unstable with respect to production of
$e^+e^-$-pairs. On account of the Pauli principle the number of produced pairs is
determined  by the number of discrete levels, which have descended into the lower
continuum. Passing through the Coulomb barrier positrons  go out to infinity, while
electrons remain near the nucleus, partially screening its charge. Thus, a naked nucleus
of supercritical charge $Z
> Z_{cr}$ will envelop itself with an electron shell created out of the vacuum; we can
call this shell ``the vacuum shell''

If $Z\alpha \gg 1$, the vacuum shell contains many electrons \footnote{The values of
critical charge  $Z_{cr}$ for highly excited atomic states were calculated in ref.[16].}
and statistical approach is necessary. The relativistic Thomas-Fermi equation [7,18,19]
can be applied to calculate electron density $n_e(r)$. Let $V(r)$ be the self-consistent
potential for an electron, taking into account both the field of the nucleus and the
average field created by other electrons of the vacuum shell. In WKB-approximation the
electron momentum  is
\be p(r) = [~(\varepsilon - V(r))^2 -m^2_e~]^{1/2},~~~ \hbar = c =1 \ee   (in the WKB
formula (10) the spin of electron is neglected, which is valid  for large $Z \gg 137$).
The vacuum shell of super-heavy nucleus is degenerated relativistic Fermi-gas with
electron density
\be n_e(r) = \frac{P_{max}^3}{3\pi^2} = \frac{1}{3\pi^2} (V^2 + 2 m_e V)^{3/2},\ee where
 the value of  $P_{max}$ follows from eq.(10) at $\varepsilon=-m_e$, since we are
interested only in the electrons that have dived into the continuum of the negative
energy states. The spatial distribution of vacuum electrons is determined by the
relativistic Thomas-Fermi equation
\be \Delta V = -4\pi e^2 \left\{ \frac{1}{3\pi^2} (V^2 +2m_e V)^{3/2} - n_p(r)\right
\}\ee with the boundary  conditions: $V(\infty)=0$ (due to global charge neutrality  of
the system) and finiteness of $V(0)$. Here $n_p(r) = n_p ~\theta(R_c -r)$ is the proton
density, $n_p = N_p ~n_0/A\approx$\\ $ 0.25$ $m^3_{\pi},~n_0 = 3A/4\pi R^3_c$ is the
ordinary nuclear density, $N_p \equiv Z$ is the number of protons and $R_c = N^{1/3}_p
m^{-1}_{\pi}$ is the core radius.

The density $n_e(r)$ of electrons is determined also by the  Fermi energy condition on
their Fermi momentum, $P_e^F = P_{max}$:
\be E^F_e = [~(P_e^F ) + m^2_e~]^{1/2} - m_e - V(r) =0,\ee which immediately leads to
eq.(10). The equations  for neutron, proton and electron densities have been integrated
numerically [20].

If $N_p e^3 >> 1$, the electric  field is concentrated in a narrow transition layer
[7,18] of thickness $\sim \lambda\approx 15$ fm near $r=R_c >> \lambda$, therefore
geometry reduces to the plane one. In the variables $\chi$ and $\xi$ one has
\footnote{Note that the thickness  of the transition layer $\lambda$ does not depend on
values of the core radius $R_c$ and the mass number $A$, if $R_c \gg \lambda$.}
$$ V(r) =-(3\pi^2 n_p)^{1/3}\chi,~~~\xi =  (r-R_c)/\lambda, $$
\be \lambda^{-1} = 2(\pi/3)^{1/6} \sqrt{\alpha}~ n_p^{1/3} \approx \sqrt{\alpha}~ m_{\pi},\ee where $\lambda\ll l_C$:

$$\lambda /l_C \sim \frac{1}{\sqrt{\alpha}} \cdot \frac{m_e}{m_{\pi}} \approx
\frac{1}{25}.\eqno{(14')}$$

Therefore  Eq.(12) becomes
\be d^2 \chi/d\xi^2 =\chi^3 -\theta(-\xi),~~~\chi(-\infty)=1,~~~\chi(\infty) =0\ee and
can be solved analitically [18]:
\be \chi(\xi) = \left\{ \begin{array}{ll} 1-3[~1+2^{-1/2}
\mbox{sinh}(a-\xi\sqrt{3})~]^{-1}, & \xi < 0,\\ & \\ 2^{1/2}(\xi +b)^{-1}, & \xi >
0,\end{array} \right.\ee where $\theta(x)$ is the Heaviside step function and the
integration constants are
$$ a = \mbox{Arsh} (11\sqrt{2}) = 3.439,~~~b = \frac{4}{3} \sqrt{2} =
1.868.\eqno(16')$$ Note that at $\xi < 0$, i.e. inside the superheavy nucleus
\be \chi(\xi) \approx  1- 0.272 ~\exp (\xi \sqrt{3}) \to 1, ~~~\mid \xi \mid \gg 1\ee

The electric field of the system
\be E(\xi) = \biggl ( \frac{3^5 \pi}{4}\biggr )^{1/6} \sqrt{\alpha} ~\frac{m^2_{\pi}}{e}
\chi'(\xi)\ee is damped exponentially  inside the nucleus: $ E(\xi) \propto \exp
(\xi\sqrt{3})~\mbox{as}~\xi \to \infty$ and $E(\xi) \propto \xi^{-2}\to 0$ in the outer
 region, $ r > R_c$. The field attains its maximal strength at the edge of super-charged
nucleus
\be E_{max} = 0.95 \sqrt{\alpha}~ \frac{m_{\pi}^2}{e} \approx \sqrt{\alpha} \biggl
(\frac{m_{\pi}}{m_e}\biggr )^2 E_{cr}, ~~~r=R_c.\ee So, $E_{max} \approx 6000~ E_{cr}$,
which exceeds the characteristic field (6) in QED and is of the same order of magnitude
 as electric field (7) at surfaces of  heavy nuclei with $Z\alpha \ga 1$.

In the region $R_c - r \gg \lambda$ the electric field is practically absent, and the
electrically neutral plasma is formed inside the supercharged  nucleus, where  the
densities $n_e$ and $n_p$ are equal and  the potential is practically constant:
\be V(r) \approx V(0) = -\biggl (\frac{9\pi}{4}\biggr )^{1/3} m_{\pi}\approx -1.92~
m_{\pi}.\ee The uncompensated charge is situated in a layer of finite thickness $\sim
\lambda$ near the  edge of the nucleus, $\hbar/m_{\pi} c \ll \lambda \ll \hbar/m_e c$.
Though the formation of electrically neutral plasma inside a supercharged nucleus, $Ze^3
\gg 1$, strongly diminishes the Coulomb energy of nucleus \footnote{Due to the screening
effect, the Coulomb energy ${\cal{E}}^{(0)}_C = 3(N_p e)^2 /5 R_c $ of a uniformly
charged sphere (without screening) diminishes by 1.7 $Ze^3$ times of magnitude  [16]. },
but it remains  positive and impedes the stability of such gigantic nuclei. So, the
conclusion of ref.[18] is that nuclei with a mass number $A\sim 10^4 - 10^6$ are unstable
due to the Coulomb repulsion of protons and can not exist in Nature.

However, the situation may be changed considerably if one  accounts a gravitational
attraction. Let us start with a simple qualitative  estimate. The Coulomb energy is
${\cal{E}}_C \sim E_{max}^2 R^2_c\lambda$, which is  mainly distributed within a thin
shell of  width $\lambda$ and radius $R_c \gg \lambda$. To ensure the stability of the
system, the attractive gravitational energy of the shell (its mass $m\sim
M\lambda/R_c,~M=A m_n$ is mass of the core)

\be {\cal{E}}_{gr} \approx - \frac{G Mm}{R_c} \sim -\frac{G M^2\lambda}{R^2_c} = -
\frac{Gm^2_nA^2 \lambda}{R^2_c}\ee  has to be larger than the repulsive Coulomb energy
${\cal{E}}_C$. Since $\mid {\cal{E}}_{gr}\mid \sim A^{4/3}$, while ${\cal{E}}_C \sim
A^{2/3}$ as $A \to \infty$, a crossing $\mid {\cal{E}}_{gr}\mid = {\cal{E}}_C$
necessarily exists:
\be \frac{\mid {\cal{E}}_{gr}\mid}{{\cal{E}}_C} \sim G m^2_n A^{2/3} =(m_n/m_{Planck})^2
A^{2/3},\ee where $m_{Planck}=\sqrt{\hbar c/G}\sim 10^{-5} g$ is the Planck mass and
$m_n\sim 10^{-24} g$ is the nucleon mass. So, $\mid {\cal{E}}_{gr}\mid > {\cal{E}}_C$ at
\be A \ga (m_{Planck}/m_n)^3\approx 10^{57},~~~R_c =1.2 A^{1/3}~\mbox{fm}~\sim 10^6
\mbox{cm},\ee
 which are typical values for neutron stars. The Coulomb
repulsion of protons, screened by relativistic electrons, is now balanced by
gravitational  forces.

The more accurate derivation of gravitational and electrodynamical stability is based on
the analytic solution (16) of the Thomas-Fermi equation. The Coulomb energy ${\cal{E}}_C$
and gravitational energy ${\cal{E}}_{gr}$ of the thin proton shell are [29]

$${\cal{E}}_C =\int\limits \frac{E^2}{8\pi} d^3 r = \frac{R^3_c~(e V(0))^3} {(3\pi
\alpha)^{1/2}} \int\limits^{\infty}_{-\infty} [~\chi'(\xi)~]^2 d\xi =$$
\be =\frac{0.345}{\sqrt{\alpha}} A^{2/3} \biggl ( \frac{N_p}{A}\biggr )^{2/3} m_{\pi} c^2,\ee
\be {\cal{E}}_{gr} = - \frac{G M m}{R_c} \approx \frac{-3~ G m^2_n}{\sqrt{\alpha}}
A^{4/3}\biggl (\frac{N_p}{A}\biggr )^{1/3} \frac{m_{\pi}e}{\hbar},\ee where $m$ is the
mass of the layer and   $Gm^2_n =\hbar c~ (m_n/m_{Planck})^2$. Hence,
$\mid{\cal{E}}_{gr}\mid
> {\cal{E}}_C$ at $A > A_R$, \be A_R \approx 0,039\biggl (\frac{N_p}{A}\biggr )^{1/2} \biggl (
\frac{m_{Planck}}{m_n}\biggr )^3,\ee which establishes a lower limit for the mass number
$A$ necessary for the stability of the massive nuclear density cores.

However, besides the Coulomb energy ${\cal{E}}_C$, the kinetic energy of the degenerated
electronic Fermi-gas exists, \be {\cal{E}}_{kin} = n_e \langle p \rangle V =\frac{3}{4}
N_p P_F,\ee where we took into account that the mean energy of particles in the
degenerated relativistic Fermi-gas is $\langle \varepsilon \rangle = \langle p \rangle
=\frac{3}{4} P_F$. The energy ${\cal{E}}_{kin}$ also impedes the stability of the system
and it should be compensated by the total gravitational energy of the core ${\cal{E}}_G =
-3G M^2/5R_c$. A simple calculation shows that $\mid {\cal{E}}_G\mid > {\cal{E}}_{kin}$
for
\be A > c_1 \biggl ( \frac{N_p}{A}\biggr )^2 \biggl ( \frac{m_{Planck}}{m_n}\biggr )^3,
\ee where $c_1$ is a numerical constant of the order of unity. So, we again arrive at the
condition, similar to Eq.(26), which is necessary  for stability of massive nuclear
density cores.  Therefore it seems possible to formulate a consistent stable model of
massive cores in terms of gravitational, strong, electromagnetic and weak interactions
and quantum statistics. Certainly, many aspects of the problem of stability remain
unsolved and further investigations are necessary.

This work is based  on the papers [18,29]. I would like to thank Professor R.Ruffini, who
initiated this work, and also V.Mur, L.Okun', M.Trusov, G.Vereshchagin, D.Voskresensky,
and S.-S.Xue for valuable discussions and remarks. The work was partially supported by
ICRANet (Pescara, Italy) and Russian Foundation for Fundamental Research, project
07-02-01116.


\begin{thebibliography}{99}

\bibitem{1} W.Gordon, Zeits. Phys. {\bf 48}, 11 (1928); {\bf 49}, 180 (1928).
\bibitem{2} C.G.Darwin, Proc. Roy. Soc London {\bf A118}, 654 (1928).
\bibitem{3} P.A.M.Dirac, {\it Principles of Quantum Mechanics}, Clarendon Press, Oxford
(1958).
\bibitem{4} I.Pomeranchuk and Ya.Smorodinsky, J.Phys.USSR {\bf 9}, 97 (1945).
\bibitem{5} W.Pieper and W.Greiner, Zeits. Phys. {\bf 218}, 327 (1969).
\bibitem{6} B.M\"uller, H.Peitz, J.Rafelski, and W.Greiner, Phys.Rev.Lett. {\bf 32}, 324
(1974).
\bibitem{7} B.M\"uller and J.Rafelski, Phys.Rev.Let. {\bf 34}, 349 (1975).
\bibitem{8} J.Rafelski, B.M\"uller, and W.Greiner, Nucl.Phys. {\bf B68}, 585 (1974).
\bibitem{9} J.Rafelski, L.P.Fulcher, and A.Klein, Phys.Rep. {\bf 38C}, 227 (1978).
\bibitem{10} W.Greiner, B.M\"uller, and J.Rafelski, {\it Quantum Electrodynamics of Strong
Fields}, Springer, Berlin (1985).
\bibitem{11} V.S.Popov, JETP Lett. {\bf 11}, 162 (1970); Sov.J.Nucl.Phys. {\bf 12}, 235
(1971).
\bibitem{12} V.S.Popov, JETP {\bf 32}, 526 (1971); {\bf 33}, 665 (1971).
\bibitem{13} Ya.B.Zel'dovich and V.S.Popov, Uspekhi Fiz.Nauk {\bf 105}, 403 (1971);
Sov.Phys.Usp.

{\bf 14}, 673 (1972).
\bibitem{14} S.S.Gershtein and V.S.Popov,  Lett. Nuovo Cim. {\bf 6}, 593 (1973).
\bibitem{16} S.J.Brodsky, Comm.Atom.Molec.Phys. {\bf 4}, 109 (1974).
\bibitem{15} V.L.Eletsky and V.S.Popov, Sov.J.Nucl.Phys. {\bf 25}, 587 (1977).
\bibitem{17} L.Okun, Comm.Nucl.Part.Phys. {\bf 6}, 25 (1974).
\bibitem{18} A.B.Migdal, D.N.Voskresensky, and V.S.Popov, JETP Lett. {\bf 24}, 163 (1976);
Sov.Phys.JETP {\bf 45}, 436 (1977).
\bibitem{19a} R.Ruffini and L.Stella, Phys.Lett. {\bf B102}, 442 (1981).
\bibitem{19} R.Ruffini, M.Rotondo, and S.-S.Xue, Int.J.Mod.Phys.{\bf D16}, 1 (2007).
\bibitem{20} V.S.Popov, Phys.Atomic Nuclei {\bf 64}, 367 (2001).
\bibitem{21} J.Schwinger, Phys.Rev. {\bf 82}, 664 (1951).
\bibitem{22} E.Brezin and C.Itzykson, Phys.Rev. {\bf D2}, 1191 (1970).
\bibitem{23} V.S.Popov, JETP Lett. {\bf 13}, 185 (1971), {\bf 18}, 255 (1973);
 Sov.Phys.JETP {\bf 34}, 709 (1972).
\bibitem{24} A.Ringwald, Phys.Lett. {\bf B510}, 107 (2001).
\bibitem{25} V.S.Popov, JETP Lett. {\bf 74}, 133 (2001); Phys. Lett. {\bf A298}, 83
(2002); JETP {\bf 94}, 1057 (2002).
\bibitem{26} N.B.Narozhny, S.S.Bulanov, V.D.Mur, and V.S.Popov, Phys.Lett. {\bf A330}, 1
(2004);
JETP {\bf 102}, 9 (2006).
\bibitem{27} R.Ruffini, G.V. Vereshchagin and S.-S.Xue, Physics Reports {\bf 487}, 1 (2010).
\bibitem{28} V.Popov, M.Rotondo, R.Ruffini, and S.-S.Xue, in preparation (2009).
\bibitem{30} L.D.Landau and E.M.Lifshitz, {\it Non-Relativistic Quantum Mechanics},
Pergamon Press, Oxford, 1975.
\bibitem{31} K.M.Case, Phys.Rev. {\bf 80}, 797 (1950).

\end{thebibliography}
\end{document}